\documentstyle[floats,prd,aps,epsf,eqsecnum,12pt]{revtex}
\makeatletter
\newbox\tempboxa
\newdimen\captionboxsubcount 
\def\capsize#1{\captionboxsubcount=#1pt}
\newdimen\captionboxsub
\captionboxsub=\hsize \advance\captionboxsub by -\captionboxsubcount
\advance\captionboxsub by -\captionboxsubcount
\long\def\@makecaption#1#2{
 \setbox\@tempboxa\hbox{#1 #2}
 \ifdim \wd\@tempboxa >\captionboxsub 
\rightskip=\captionboxsubcount \leftskip=\captionboxsubcount #1 #2 
\else \hbox to\hsize{\hfil\box\@tempboxa\hfil} 
 \fi}
\makeatother
\capsize{30}

\begin{document}
\begin{titlepage}

\begin{flushright}
\begin{minipage}{5cm}
\begin{flushleft}
\small
\baselineskip = 13pt
YCTP-P19-98\\
hep-th/9806229 \\
\end{flushleft}
\end{minipage}
\end{flushright}

\begin{center}
\Large\bf
From Super QCD to QCD
\end{center}

\vskip 1cm

\footnotesep = 12pt

\begin{center}
\large
 Francesco {\sc Sannino}
\footnote{Electronic address : 
{\tt sannino@apocalypse.physics.yale.edu}}\\ 
\vskip .6cm
{\it Department of Physics, Yale University, New Haven, 
CT 06520-8120, USA.}
\end{center}

\vfill
\begin{center}
\bf
Abstract
\end{center}
\begin{abstract}
\baselineskip = 17pt
We  present a ``toy'' model for breaking 
supersymmetric gauge theories at the effective Lagrangian level. 
We show that it is possible to achieve the decoupling of gluinos 
and squarks, below a given supersymmetry breaking scale $m$, 
in the fundamental theory for super QCD once a suitable 
choice of supersymmetry breaking terms is made.  
A key feature of the model is the description of the ordinary QCD 
degrees of freedom via the auxiliary fields of the supersymmetric 
effective Lagrangian. 
Once the anomaly induced effective QCD meson 
potential is deduced we also suggest a decoupling procedure, when a flavored 
quark becomes massive, which mimics the one employed by Seiberg for 
supersymmetric theories. It is seen that, after quark decoupling, 
the QCD potential naturally converts to the one with one less flavor. 
Finally we investigate the $N_c$ and $N_f$ dependence of the $\eta^{\prime}$ 
mass. 
\end{abstract}

\vfill

\end{titlepage}

\setcounter{footnote}{0}

\section*{General Strategy}
In the last few years there has been an enormous progress in understanding 
 supersymmetric gauge theories via effective Lagrangians. Such a progress 
is partially due to some papers of Seiberg \cite{Seiberg} and Seiberg and 
Witten \cite{Seiberg-Witten} in which a number of ``exact results'' were 
obtained. There are already several review articles 
\cite{IntSeiberg,Peskin,DiVecchia}. 

It is natural to expect that information obtained from the more 
highly constrained supersymmetric gauge theories can be used 
to learn more about ordinary gauge theories.  
Here we illustrate the general strategy behind a ``toy'' model presented in 
\cite{toy} for breaking super symmetric 
 gauge theories at the effective Lagrangian level. 

Let us consider an ``exact'' effective super potential $W$ which  
can be constructed for a given, confining, 
supersymmetric gauge theory. The superpotential, by construction, correctly 
saturates all the supersymmetric quantum anomalies. 
 The contribution to the bosonic part of the potential, contained 
in the superpotential, before imposing 
the equation of motion for the auxiliary fields, is:
\begin{equation}
-V_0\left({\cal F},{\cal \phi}\right)=\int d^2\theta \, \
W\left({\cal S},{\cal T}\right) + {\rm H.c.}\ ,
\label{potential0}
\end{equation}
where the chiral superfields ${\cal S}$ and ${\cal T}$ 
schematically describe gauge invariant supersymmetric 
composite operators whose bosonic components (${\cal \phi}$)
 respectively contain gluino-ball and squark-antisquark mesons. 
${\cal F}$  are the set of auxiliary fields 
associated with the chiral superfields ${\cal S}$ and ${\cal T}$. 
We note that the previous potential term, due to supersymmetry, is 
holomorphic in the fields, i.e. 
${V_0\left({\cal F},{\cal \phi}\right)}=
\chi\left({\cal F}, {\cal \phi}\right) + 
\chi^{\dagger}\left({\cal F}^{*}, {\cal \phi}^{*}\right)$ and $\chi$ is 
a function of the complex fields ${\cal F}$ and ${\cal \phi}$. 
The composite operators ${\cal F}$ are seen to describe the
 ordinary glue-ball and mesonic objects (see for example 
Eq.~(\ref{components})). 

 Let us imagine to add SUSY breaking terms in the 
fundamental Lagrangian whose order parameters (i.e. gluino mass and 
squark mass terms) can be {\it schematically} represented by ${\cal M}$. 
This will induce in the low energy effective theory a SUSY 
breaking potential $V_B\left({\cal M},{\cal \phi}\right)$ 
which should be added to the one in 
Eq.~(\ref{potential0}). 
The full potential for finite ${\cal M}$ can then be written as 
\begin{equation}
V\left({\cal F}, {\cal \phi}, {\cal M}\right)=
V_0\left({\cal F},{\cal \phi}\right) + 
V_B\left({\cal M},{\cal \phi}\right) \ . 
\label{potential}
\end{equation}

If ${\cal M}\ll \Lambda_S$, where $\Lambda_S$ is the SUSY 
invariant scale of the theory we are close to the supersymmetric 
limit and ${\cal F}$ must be eliminated via its equation of motion
\begin{equation}
\frac{\partial V}{\partial {\cal F}}=0 \quad \quad {\rm for}
\quad \quad {\cal M}\ll {\Lambda_S} \ .   
\end{equation}
 In the absence of the K\"ahler term, the previous 
equation simply reproduces the supersymmetric vacuum solution for 
the bosonic fields ${\cal \phi}$. To recover the 
``soft'' SUSY breaking effects \cite{Masiero-Veneziano}, 
beside modelling $V_B$ via 
supersymmetric spurions, one must consider a model for the 
K\"ahler term which, in turn, should be invariant under the anomalous 
transformations. It is amusing to note that K\"ahler 
terms in the effective theory can be regarded as higher order in 
a derivative expansion with respect to the invariant 
scale of the theory and that 
${\cal M}/\Lambda_S$ corrections arise when also K\"ahler terms are 
present. 

If ${\cal M}\gg \Lambda_S$ the light degrees of freedoms are now 
the ordinary fields (quarks, gluons, etc.). These seems to be 
contained in the auxiliary fields of the effective Lagrangian description. 
Hence we expect ${\cal F}$ to remain, while ${\cal \phi}$, which describes 
the colorless objects made of gluinos and squarks, to decouple. 
This can be obtained by assuming, as proposed in \cite{toy}, 
the following equation of motion
\begin{equation}
\frac{\partial V}{\partial {\cal \phi}}=0 \quad \quad {\rm for}
\quad \quad {\cal M}\gg {\Lambda_S} \ .
\label{eqphi}
\end{equation} 
This provides the relation 
\begin{equation}
{\cal \phi}={\cal \phi}\left({\cal F},{\cal M},{\Lambda_S}\right) \ ,
\end{equation}
By substituting the previous expression in Eq.~(\ref{potential}) we 
obtain the potential: 
\begin{equation}
V=V\left({\cal F},{\cal M},{\Lambda_S}\right) \ .
\end{equation} 
A smooth decoupling is achieved if, at least at one loop in 
the underlying theories, the scales 
${\cal M}$ and ${\Lambda_S}$ combine 
in the unique scale $\Lambda$ associated with the ordinary gauge theory.

Of course the knowledge of $V_B$ is essential. 
We partially fix the breaking potential \cite{toy} by requiring the anomalies 
(trace anomaly as well as global anomalies) to match at the one loop level 
and assuming holomorphy for the breaking potential. 
The latter assumption is partially supported by the holomorphic 
behavior of the one--loop QCD coupling constant.  

\section*{From Super Yang Mills to Yang Mills}
\label{sec:SUSYYANGMILLS}

In this section we illustrate, in some detail, how the previous 
strategy works in the SUSY Yang Mills case \cite{toy}.  
The effective Lagrangian for Super Yang Mills 
was given \cite{refVY} by Veneziano 
and Yankielowicz (VY) and is described by the Lagrangian 
\begin{equation}
{\cal L} = \frac{9}{\alpha} \int d^2\!\theta d^2\!{\bar{\theta}} 
\left(S S^{\dagger}\right)^{\frac{1}{3}} + 
\left\{\int d^2\theta\,S\left[{\rm ln}\left(\frac{S}{\Lambda_{SYM}^3}
\right)^{N_c} - 
N_c \right] + {\rm H.c.} \right\} \ ,
\label{VY}
\end{equation}
 where $\Lambda_{SYM}$ is the super 
$SU(N_c)$ Yang Mills invariant scale and the 
chiral superfield $S$ stands for the composite object 
$S=\frac{g^2}{32\pi^2}W^{\alpha}_{a}W_{\alpha a}$. Here $g$ is the gauge 
coupling constant and $W^{\alpha}_{a}$ is the supersymmetric field strength. 
At the component level with 
 $S(y)=\phi(y) + \sqrt{2} \theta \psi(y) + \theta^2 F(y)$ we have
\begin{equation} 
\phi\approx \lambda^2\ , \quad \psi \approx \sigma^{mn} \lambda_a F_{mn,a}
\quad  {\rm and} \quad F \approx -\frac{1}{2} F^{mn}_a F_{mn,a} - 
\frac{i}{4} \epsilon_{mnrs}F^{mn}_a F^{rs}_ a \ .
\label{components}
\end{equation}
$\lambda^{\alpha}_a$ is the gluino field, $F_{mn,a}$ the gauge field 
strength.

We interpret the complex field $\phi$ as representing scalar and 
pseudoscalar gluino balls while $\psi$ is their fermionic partner. 
The auxiliary field $F$ is seen to contain scalar and pseudoscalar glueball 
type objects. 

 Equation (\ref{VY}) describes the vacuum 
of the theory and saturates 
the anomalous Ward identities at tree level. These anomalies arise 
in the axial current of the gluino field, the trace of the energy momentum 
tensor and in the special superconformal current. In supersymmetry these 
three anomalies belong to the same supermultiplet and hence 
are not independent. For example
\begin{eqnarray}
\theta^m_m&=&3\,N_c\,\left(F + F^{*}\right) = 
-\frac{3 N_c g^2}{32\pi^2} F^{mn}_a F_{mn,a} \ , \label{trace}\\
\partial^m J_{m}^5 &=&2i\, N_c \, \left(F-F^{*}\right)= 
\frac{N_c g^2}{32\pi^2} \epsilon_{mnrs}F^{mn}_a F^{rs}_a \ .
\label{axial}
\end{eqnarray}
where $J^5_m = \bar{\lambda}_a \bar{\sigma}_m \lambda_a$ is the axial current.
The effective Lagrangian yields \cite{refVY} the 
gluino condensation of the form $
\langle\phi\rangle=-\frac{g^2}{32\pi^2} \langle\lambda^2\rangle = 
\Lambda^3 e^{\frac{2\pi i k}{N_c}}$ where $k=0,1,2,\cdots,(N_c -1)$.

Masiero and Veneziano investigated the ``soft'' supersymmetry 
breaking regime
\cite{Masiero-Veneziano} 
by introducing a ``gluino mass term'' in the Lagrangian
\begin{equation}
{\cal L}=\cdots + m \left(\phi + \phi^*\right) \ ,
\label{soft}
\end{equation}
with the softness restriction $m\ll \Lambda_{SYM}$. The results 
of this model \cite{Masiero-Veneziano} indicate that 
the theory is ``trying'' to approach the ordinary Yang Mills case. 

It seems very desirable to extend this model to the case of large 
$m(\gg\Lambda_{SYM})$ in which the superparticles decouple from the 
theory and the theory gets reexpressed in terms of ordinary glueball 
fields.  In Ref.~\cite{toy}
we proposed a toy model which accomplishes these goals. 
Our approach is based on the general strategy presented in the previous 
paragraph which we specialize, here, for the super Yang Mills case:
\begin{itemize}
\item[i)]{We shall concentrate completely on the 
superpotential. This 
contains all the information on the anomaly structure and seems to be 
the least model dependent part of the effective Lagrangian.}
\item[ii)]{ 
We will show that the generalisation of the supersymmetry breaking 
term Eq.~(\ref{soft}) to 
\begin{equation}
V_B=- m^{\delta} \phi^{\gamma} + {\rm H.c.} \ ,
\label{breaking}
\end{equation}
where $\delta=4-3\gamma$ and 
${\gamma=\frac{12}{11}}$ automatically accomplishes 
the decoupling of the underlying gluino degree of freedom at the scale $m$. 
The deviation of the exponent $\gamma$ from unity is being thought of as 
an effective description of the evolution of the symmetry breaker 
Eq.~(\ref{soft}) for large $m$.}
\item[iii)]{Since the Yang Mills 
fields of interest are contained in $F$ 
the heavy gluino ball field $\phi$ is eliminated 
by its equation of motion ${\frac{\partial V}{\partial \phi}=0}$ 
(see Eq.~(\ref{eqphi})).}
\end{itemize}

The potential of our model 
\begin{equation}
V(F,\phi)=-F\,{\rm ln}\left(\frac{\phi}{\Lambda_{SYM}^3}\right)^{N_c} - 
m^{\delta} \phi^{\gamma} + {\rm H.c.} \ ,
\label{totpotentialVY}
\end{equation}
provides the equation of motion, 
 $\frac{\partial V}{\partial \phi}=0$ for eliminating $\phi$:
$
\phi^{\gamma}=-\frac{N_c F}{\gamma m^{\delta}}\label{eomVY}$. 
Our physical requirement is that the presence of the symmetry breaker 
Eq.~(\ref{breaking}) should convert the anomalous quantity $\theta^m_m$ 
into the appropriate one for the ordinary Yang Mills theory. 
This is in the same spirit as the well known \cite{Wittendec} criterion for 
decoupling a heavy flavor (at the one loop level) in QCD. 
We compute 
$\theta^m_m$ at tree level obtaining
\begin{equation}
\theta^m_m=\frac{4N_c}{\gamma}\left(F + F^*\right)= -
\frac{4N_c}{\gamma} \left(\frac{g^2}{32\pi^2}F^{mn}_aF_{mn,a}\right) \ .
\label{stYM}
\end{equation}
Now the 1--loop anomaly in the underlying theory is given by
\begin{equation}
\theta^m_m=-b\frac{g^2}{32\pi^2}F^{mn}_aF_{mn,a} \ ,
\label{stunder}
\end{equation}
where $b=3N_c$ for supersymmetric Yang Mills and $b=\frac{11}{3}N_c$ for 
ordinary Yang Mills. In order that Eq.~(\ref{stYM}) match Eq.~(\ref{stunder}) 
for ordinary Yang Mills we evidently require 
 $\gamma=\frac{12}{11}$ as mentioned above. With $\phi$ 
eliminated in terms of $F$ the potential becomes
\begin{equation}
V(F)=-\frac{11N_c}{12}F\left[{\rm ln}\left(\frac{-11 N_c F}
{12 m^{\frac{8}{11}} 
\Lambda_{SYM}^{\frac{36}{11}}}\right) - 1 \right] + {\rm H.c.} \ .
\label{YM}
\end{equation}
We now check that this is consistent with a physical picture in which 
the gauge coupling constant evolves according to the super Yang Mills 
beta--function above scale $m$ and according to the Yang Mills beta--function 
below scale $m$. Since the coupling constant at scale $\mu$ is given 
by ${\left(\frac{\Lambda_{SYM}}{\mu}\right)^b = 
\exp\left(\frac{-8\pi^2}{g^2(\mu)}\right)}$, the matching at $\mu=m$ requires 
${\left(\frac{\Lambda_{SYM}}{m}\right)^b=
\left(\frac{\Lambda_{YM}}{m}\right)^{b_{YM}}}$, which yields
$
\Lambda_{YM}^4=m^{\frac{8}{11}} \Lambda_{SYM}^{\frac{36}{11}} \ ,
\label{matchVY}$ 
in agreement with the combination appearing in Eq.~(\ref{YM}).
The Lagrangian in Eq.~(\ref{YM}) manifestly depends only on quantities 
associated with the Yang Mills theory, the gluino degree of freedom 
having been consistently decoupled. Equation~(\ref{YM}) thus seems to be 
a reasonable candidate for the potential term of a model describing 
the trace anomaly in Yang Mills theory. 

The model is seen to contain a scalar 
glueball field ${\rm Re}F$ and a pseudoscalar glueball field 
${\rm Im}F$. In order to relate our present results to previous 
investigations \footnote{See discussion in Ref.~\cite{toy}} 
we also eliminate 
${\rm Im}F$ by its equation of motion which yields ${\rm Im}F=0$. 
Substituting this back into 
Eq.~(\ref{YM}) and 
using the notation 
$H=\frac{11N_c }{3}\frac{g^2}{32\pi^2}F^{mn}_aF_{mn,a}\label{Hdef}$, 
leads to the potential function
\begin{equation}
V(H)=\frac{H}{4}\, {\rm ln} \left(\frac{H}{8e\Lambda^4_{YM}}\right) \ .
\label{YMH}
\end{equation}
This may be considered as a zeroth order model 
\cite{joe,MS,SST} 
for Yang Mills theory 
in which the only field present is a scalar glueball. $V(H)$ has 
a minimum at $\langle H\rangle = 8\Lambda^4_{YM}$, at which point 
$\langle V\rangle=-2 \Lambda^4_{YM}$. This corresponds to a 
magnetic--type condensation of the glueball 
field $H$. A number of phenomenological questions have been discussed 
using toy models based on Eq.~(\ref{YMH}) \cite{SST,GJJS,GJJS2}.  

It can also be shown that $m_{\psi}\rightarrow \infty$ in the case
 $m\rightarrow \infty$; thus, as expected,  ${\psi}$ decouples \cite{toy}.

\section*{From Super QCD to QCD and Quark Decoupling}

The more complicated case of adding matter fields with number of flavors $N_f$ 
less than number of colors $N_c$ with $N_c\ne2$ has been analyzed in 
\cite{toy,HSS}. Here we summarize some of the relevant results. The needed 
``mesonic'' composite superfield is the complex $N_f\times N_f$ matrix 
$T_{ij}=Q_i \tilde{Q_j}=t_{ij}+\sqrt{2}\theta\psi_{T{ij}}+\theta^2 M_{ij}$, 
where $i$ and $j$ are flavor indices. $Q$ and $\tilde{Q}$ 
are the quark anti-quark chiral superfields. It can be seen that the 
mesonic auxiliary field $M\approx - \psi_Q \psi_{\tilde{Q}}$ 
contains the ordinary quark-antiquark meson field while 
$t=\phi_Q \phi_{\tilde{Q}}$ describes the squark anti-squark composite 
operator. In Ref.\cite{TVY} a straightforward generalization of 
the supersymmetric potential presented in Eq.~(\ref{VY}) 
for $N_f<N_c$ was derived. By a suitable decoupling of the squark as well 
as gluino degrees of freedom (see Ref.~\cite{toy} for more 
details) the following potential 
for ordinary QCD can be deduced 
\begin{equation}  
V\left(M\right)=- C\left(N_c,N_f\right)
\left[
\frac{\Lambda_{QCD}^
{\frac{11}{3}N_c - \frac{2}{3}N_f}}
{{\rm det}\, M}
\right] ^{\frac{12}{11\left(N_c - N_f \right)}} \ ,
\label{hss-potential}
\end{equation}
where $\Lambda_{QCD}$ is the invariant QCD scale and $C\left(N_c,N_f\right)$ 
is a definite positive quantity which cannot be fixed by requiring the 
potential to satisfy the anomalies. We also eliminated the glue-ball 
degrees of freedom via their equation of motion (see Ref.~\cite{HSS}).  

The potential for the meson variables in Eq.~(\ref{hss-potential}) 
is similar to the 
effective Affleck-Dine-Seiberg (ADS) \cite{refADS} superpotential for massless 
super QCD theory with $N_f<N_c$ 
\begin{equation}
W_{ADS}\left(T\right)=-\left(N_c - N_f\right)
\left[
\frac{\Lambda_S^{3N_c-N_f}}{{\rm det}\, T}
\right]^{\frac{1}{N_c - N_f}} \ ,
\label{ads-spotential}
\end{equation}
where $\Lambda_S$ is the invariant scale of SQCD.

An intriguing feature of the potential 
in Eq.~(\ref{hss-potential}) is that it presents a fall to the origin 
rather than a run-away vacuum associated with the ADS superpotential. 
The fall to the origin 
can be fixed  by adding an anomalous free non holomorphic term in 
the manner outlined in \cite{toy}, which in turn requires spontaneous 
chiral symmetry breaking. It is worth noticing \cite{HSS} that one can 
directly derive Eq.~(\ref{hss-potential}) from QCD, if one assumes, 
besides the correct anomalous transformations, also one--loop holomorphicity 
in the QCD coupling constant \cite{HSS,smrst}.    

As for the SUSY case \cite{Seiberg} we can partially deduce 
the $N_f$ and $N_c$ dependence of $C$ by 
defining a decoupling procedure for quarks. In Ref.~\cite{HSS} it has been  
shown that by adding the following, generilized quark mass operator, 
\begin{equation}
V_m=-m^{\Delta} {M^{N_f}_{N_f}}^{\Gamma} + {\rm H.c.} \ ,
\end{equation}
 to the potential in Eq.~(\ref{hss-potential}), with $\Delta=4 - 3 \Gamma$,  
is possible to obtain a complete decoupling when a flavored 
quark becomes massive. This procedures mimics the one employed by Sieberg 
for supersymmetric gauge theories. It is seen that, after 
decoupling, the QCD potential naturally converts to the one with one less 
flavor provided that $\Gamma=12/11$ and the coefficient $C$ has the following functional 
form
\begin{equation}
C\left(N_c,N_f\right)=\left(N_c - N_f \right) D\left(N_c\right) 
^{\frac{1}{N_c-N_f}}\ .
\end{equation} 
$D\left(N_c\right)$ is an unknown $N_c$ dependent function. 
It is interesting to contrast 
the coefficient of the ``holomorphic'' part of the QCD potential 
Eq.~(\ref{hss-potential}) 
with Seiberg's result \cite{Seiberg} 
$C\left(N_c,N_f\right)=N_c-N_f$ for the coefficient 
of the ADS superpotential (Eq.~(\ref{ads-spotential})). Clearly, 
in the SUSY case the analog of $D\left(N_c\right)$ 
is just a constant. This feature arises in the SUSY case because of the existence of squark 
fields which can break the gauge and flavor symmetries by the 
Higgs mechanism. The 
possibility of a non-constant $D\left(N_c\right)$ factor can 
thus be taken as an 
indication that there is no Higgs mechanism 
present in QCD-like theories. 

Finally the knowledge about $C(N_c,N_f)$ can be used to 
suggest that the well known \cite{Wittena} large $N_c$ 
behaviour of the $\eta^{\prime}$ (pseudoscalar singlet) meson 
mass should also include an $N_f$ dependence of the form:
\begin{equation}
M^2_{\eta^{\prime}}\propto \frac{N_f}{N_c - N_f} \Lambda^2 
\quad\quad 
\left(N_f < N_c\right)  \ .
\label{mass}
\end{equation}
It is amusing to observe that when $N_f$ is close to $N_c$ the 
resulting pole in 
Eq.~(\ref{mass}) suggests a possible enhancement mechanism for 
the $\eta^{\prime}$ mass. 
This would explain the unusually large value of this quantity in the 
realistic three 
flavor case.

In future we would like to understand the very important and yet elusive case 
$N_f=N_c$, where as argued in Ref.~\cite{HSS} we expect non holomorphic terms 
to be relevant, since the coefficient $C\left(N_c,N_f\right)$, in analogy 
with the SUSY case, of the anomalous potential vanishes for $N_f=N_c$.

In Ref.~\cite{KS}, we computed the one--loop effective action in 
the specific case of $N_f=N_c+1$ and $N_c=2$ while keeping only the 
auxiliary fields on the external legs, and in the presence of 
supersymmetry breaking terms. This procedure amounts 
to integrate out order by order, in a loop expansion, the scalar fields. 
It was shown how a 
non-trivial kinetic term for 
the auxiliary field naturally emerges, reinforcing our assumption 
that the latter can be associated with a physical field 
once the supersymmetric particles decouple.    

It is also very interesting to explore the $N_f>N_c$ case which might 
shed some light on the zero temperature chiral restoration and a possible 
relation with the conformal window \cite{AS}.

\acknowledgments
I am happy to thank Joseph Schechter for very helpful discussions. 
I also thank CERN Theoretical Division, where part of this work was 
completed. The present work has been partially supported by the US DOE  
under contract DE-FG-02-92ER-40704.

\end{document}